# Chromatic Confocal Tomography

Gargi Sharma, Kanwarpal Singh

This work was supported by the Max Planck Society for the Advancement of Science.

Gargi Sharma is a post doctoral fellow with the Max Planck Institute for the Science of Light, Erlangen 91058, Germany (e-mail: gargi.sharma@mpl.mpg.de).

Kanwarpal Singh is a Group Leader with the Max Planck Institute for the Science of Light, Erlangen 91058, Germany and affiliated faculty with the Department of Physics at Friedrich-Alexander-Universität Erlangen-Nürnberg, Erlangen 91058, Germany (e-mail: kanwarpal.singh@mpl.mpg.de).

***Abstract*—Confocal microscopy is the backbone of cellular research labs across the world but unfortunately, the imaging is restricted to a single plane. Chromatic confocal microscopy offers the possibility to image multiple planes simultaneously thus providing a manifold increase in the imaging speed whereas eliminating the need for z-axis scanning. Standard chromatic confocal systems have a limited imaging range of the order of a few hundreds of micrometers which limits their applications. In this work, using a single zinc selenide lens, we demonstrate a chromatic confocal tomography system that has an imaging range of 18 mm with an average lateral resolution of 2.46 µm and another system with a 1.55 mm imaging range with 0.86 µm average lateral resolution. Doing so, we have pushed the imaging range of chromatic confocal systems from micrometer to centimeter regime, an improvement of approximately two orders of magnitude. The proposed approach can be a powerful tool for confocal imaging of biological samples or surface profiling of industrial samples.**

***Index Terms*—Confocal microscopy, Chromatic confocal microscopy, Imaging systems, Microscopy.**

## I. INTRODUCTION

VISUALIZATION of the cellular and subcellular structures within the tissue has helped the biomedical community to understand and identify the very nature of these entities. Many of the times the anomaly initiates not at the surface but beneath. For such cases, cellular level, three-dimensional imaging is required for early disease diagnosis. Recent tomographic techniques provide an added advantage to microscopy, as such techniques allow one to visualize beneath the surface. High-resolution tomographic techniques like optical coherence tomography [1], confocal microscopy [2], chromatic confocal microscopy [3], etc have been developed in this direction.

Optical coherence tomography (OCT) is a relatively new technique that provides three-dimensional information of the sample based on low-coherence interferometry of light. Using OCT, one may achieve an axial and lateral resolution of a few micrometers. Even though OCT can achieve cellular level lateral resolution [4], it is maintained in the focal plane only, and as one moves above or below the focus, the resolution drops significantly. Confocal microscopy has been the technique of choice for biologists for cellular and subcellular tomography with lateral resolution better than 1 µm [5]. Standard confocal microscopy is a point scanning method, and one would need to scan X, Y, and Z-axis to acquire the complete tomographic information, which increases the scanning time significantly [6]. Spinning disc confocal microscopy [7] has increased the scanning speeds to a level that up to 1000 frames can be acquired in less than a second, such systems however are still very bulky and only used as benchtop imaging systems. To overcome this limitation, techniques like chromatic confocal microscopy [3, 8-20] have been developed. Using the chromatic dispersion property of light, one may extend the imaging range with a comparable lateral resolution to that of confocal microscopy. Recently specially fabricated metalenses [13] have been used for chromatic confocal microscopy achieving an imaging range of 33.5 µm but manufacturing of such lenses is still limited to specialized facilities. To the best of our knowledge, a maximum imaging range of 250 µm has been achieved using chromatic aberration of multiple achromatic and diffractive lenses where an average lateral resolution of around 1 µm is maintained within the imaging range [20].

An ideal 3-D tomographic imaging system should be able to provide a lateral resolution that allows cellular and subcellular level imaging at the maximum possible imaging range. Generally speaking, one may improve the resolution in the aforementioned techniques by using a high numerical aperture (NA) lens but this comes at the expense of reduced imaging range. A higher imaging range can be achieved by using a lower NA objective but then the lateral resolution is compromised. Therefore current advancements in this direction suffer from the trade-off between the imaging range and the resolution. For surface profiling, which is one of the main applications of chromatic confocal microscopy [10, 17, 21], multi-lens commercial systems are available with an imaging range of the order of a few centimeters but with poor lateral resolution. Unfortunately, we still lack a tomographic technique that has a superior lateral and axial resolution, and a large imaging range to visualize cellular and subcellular structures in biological samples.

In this work, we utilized the exceptionally high chromatic dispersion property of the zinc selenide (ZnSe) material-based lens to focus different wavelength components of a supercontinuum light source at different axial positions to develop a chromatic confocal tomographic (CCT) system.

## II. EXPERIMENTAL DESIGN AND METHODS

For a CCT system, one needs to consider mainly two aspects. First, the numerical aperture of the system should be as high as possible to achieve the best possible resolution. Secondly, the axial focal separation among different wavelength components should be increased significantly to achieve a maximum possible imaging range. However, there is a trade-off between the imaging range and the spatial resolution. The resolution can be improved using a high numerical aperture objective, but this, in turn, reduces the imaging range and vice versa. To quantify this relation, we have defined a parameter, extended depth resolution ratio (EDRR), which is the ratio of the imaging range and average resolution in that region. The higher the EDRR, the better the optical device's capability to image for a longer depth with good resolution. We have compared the average lateral resolution, average axial resolution the imaging range, and the EDRR of previously reported chromatic confocal systems in Supplementary information (Table S1).

We simulated our design using OpticStudio software (Zemax LLC, UK) to optimize the aforementioned parameters. To be able to achieve maximum axial separation between different wavelength components, the chromatic dispersion property of the material is utilized. The chromatic dispersion of a material can be estimated by the Abbe number; the lower the Abbe number, the higher is the chromatic dispersion of the material. ZnSe has an Abbe number of 8 in the visible wavelength range and is the smallest Abbe number material we could find whose lenses are commercially available. It has a transmission of more than 50% for the wavelength range of 0.55 μm to 15 μm. Using the OpticStudio software, we simulated the performance of three commercially available ZnSe lenses having a focal length of 6.35 mm (39469, Edmund Optics, UK), 12.7 mm (39471, Edmund Optics, UK), and 50.8 mm (39495, Edmund Optics, UK) at 10 μm wavelength.

It should be noted that all of these lenses are designed to focus a collimated beam and corrected for the spherical aberration at 10 μm wavelength. However, we have used these lenses in the visible and NIR wavelength range. It is feasible because of the reasonable transmission of ZnSe at these wavelengths. However, these lenses can not be used to focus a collimated visible/NIR beam without introducing significant spherical aberrations and thus degrading the lateral resolution. We circumvented this issue by using a diverging beam from the fiber tip (instead of a collimated beam) which was focused using the ZnSe lens. For a diverging beam, with a ceratin numerical aperture, at a certain wavelength, there is a fixed distance of the lens from the fiber tip for which the aspherical aberrations are minimum at the focal position. Using the simulations, we calculated the distance between the lens and the fiber tip for which the 700 nm wavelength has the minimum spherical aberrations.   For this fixed distance between the lens and the fiber tip, parameters for the other wavelengths of the spectrum, i.e. 500 nm and 950 nm were simulated. For all the simulations, the fiber NA of 0.14 was considered. From the simulated focal position for 500 nm, 700 nm, and 950 nm wavelengths, we calculated the imaging range, which is the distance between the focal positions of the 500 nm and the 950 nm wavelength. The lateral resolution is defined by the full-width-half-maxima (FWHM) of the focal spot and was optimized by optimizing the Strehl ratio (SR). We also calculated EDRR for all three lenses.

For developing the CCT device, we utilized a supercontinuum light source (SCL) (SC-Pro, YSL Photonics, China). The experimental design is shown in Fig. 1. The broadband light from the laser was steered through mirrors M1 and M2. The mirror M1 used in the set-up was a dichroic mirror (DMPL950T, Thorlabs Inc., USA), which reflected wavelengths up to 950 nm only. The higher wavelengths were transmitted through the mirror M1 and were blocked using a beam block. The light from the mirror M2 was coupled in the input arm (1) of a 75/25 single-mode optical coupler (TW670R3A2, Thorlabs Inc. USA) with a core diameter of 3.5 μm (single-mode cut-off wavelength 570 nm) using lens L1 (M-10X, Newport Corporation, USA). The diverging light at the output arm (2) of the fiber coupler was focused using the ZnSe lens at the sample position. The focussed spot size at the sample depends on the mode profile of the laser output from the fiber coupler (which can be multimode at wavelength < 570 nm) and the ZnSe lens performance at this wavelength. Ideally one should use a fiber that is single-mode at all the wavelengths to achieve the best focus for all the wavelengths but unfortunately, broadband fiber couplers based on such fiber are not commercially available yet. The optical power at the sample was measured to be 17 mW which is lower than the power used in standard laser scanning confocal microscopes [22, 23]. The scattered light from the sample was coupled back to the output arm (2) of the coupler and directed to the spectrometer using arm (3). The output arm (4) of the fiber coupler can be used to monitor the laser power and spectrum. For imaging, the beam over the sample was scanned by 0.5 mm at 100 Hz using an amplified piezoelectric actuator (APF710, Thorlabs Inc. USA) along fast imaging axis and by 0.5 mm at a speed of 0.1mm/s using a precision translation stage (MBT616D/M Thorlabs Inc., USA) along slow imaging axis. An area of 0.5 mm × 0.5 mm was scanned using the scanning stages and 512 × 512 axial scans were acquired to reconstruct the 3 D volumetric images of the sample. Each axial scan represented a depth of 1.55 mm and consisted of 2048 pixels. Within the spectrometer, different wavelength components of the light were spatially separated using a prism (PS855, Thorlabs Inc., USA) and focussed on a linear CCD camera (raL2048, Basler AG, Germany) sensor working at 51.2 kHz line rate using an objective lens (AF Nikkor 85 mm, Nikon, Japan).  The combination of the camera and the scanning module allowed us to acquire the full volumetric data (512 × 512 × 2048 pixels) in approximately 5.12 seconds. We have used a prism instead of a grating in the spectrometer because it provides higher transmission efficiency over a broad spectrum.

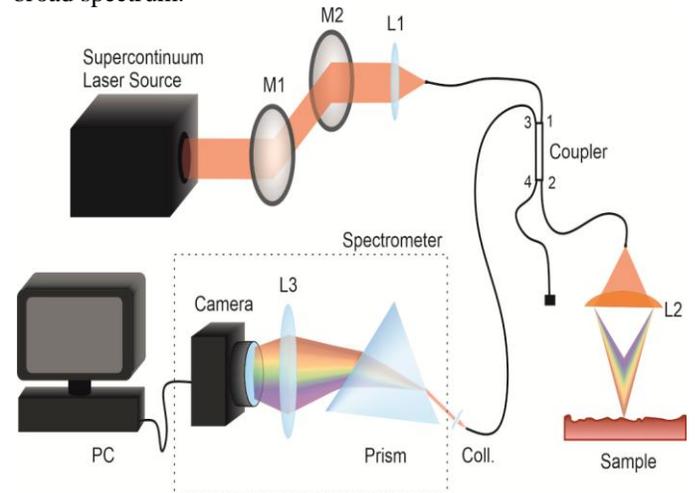

Fig. 1. Experimental set-up for the realization of the chromatic confocal tomographic device. Mirror (M), Lens (L), Personal computer (PC), Collimator (Coll).

The CCT system was tested for its performance by measuring parameters like imaging range, lateral, and axial resolution. The lateral resolution of the device was characterized by imaging the resolution target (R1L3S6P, Thorlabs Inc., USA) and calculated the focal spot size of the CCT system using the knife-edge technique [24]. The knife-edge measurement was performed by traversing the resolution target with 50 nm steps using an automated translation stage. The distance traveled by the stage between the 90% beam power transmission and 10% beam power transmission was measured and multiplied by 1.561 to acquired the focused beam diameter [24]. We also calibrated the imaging range of our system by placing a mirror at the sample position and moving it axially using a precision translation stage (MBT616D/M Thorlabs Inc., USA), and measuring the peak position of the reflected signal from the mirror, in the spectrometer as a

| Lens | $\lambda$ (nm) | Strehl Ratio | Focal position (mm) | Lateral Resolution ($\mu$m) | EDRR |
|------|------|------|------|------|------|
| 39469 (f=6.3 5 mm) | 500 nm | 0.16 | 5.5 | 0.7 | |
| | 700 nm | 0.99 | 6.57 | 1.08 | |
| | 950 nm | 0.86 | 7.07 | 1.57 | |
| | | | IR= 1.57 | Avg (1.11) | 1414 |
| 39471 (f=12. 7 mm) | 500 nm | 0.1 | 12.4 | 0.8 | |
| | 700 nm | 0.94 | 14.9 | 1.4 | |
| | 950 nm | 0.81 | 15.8 | 2.1 | |
| | | | IR=3.4 | Avg (1.43) | 2377 |
| 39495 (f=50. 8 mm) | 500 nm | 0.4 | 64.4 | 1.5 | |
| | 700 nm | 1.0 | 78.2 | 2.75 | |
| | 950 nm | 0.85 | 83.78 | 3.2 | |
| | | | IR=19.38 | Avg (2.48) | 7814 |

Simulations for the comparison of different focal length ZnSe focal length (f) lenses for the Strehl ratio, imaging range (IR), average focal spot size, and EDRR.

function of the stage position. To measure the axial resolution, we measured the full-width half maxima of the axial point spread function which was obtained by placing a mirror as a sample at different depths. To demonstrate the 3D imaging capabilities of our device with high resolution over the imaging range, we performed volumetric imaging of a nanoparticle phantom (OCTPHANTOMS.ORG, UK) with particle size < 1 $\mu$m. Further, we measured the point spread function of the nanoparticles at different depths. The potential of the device for imaging the biological samples was demonstrated by imaging the microscopic features in a swine cornea sample in 3D.

Furthermore, since the chromatic confocal microscopes have been extensively used in the industry for the height profiling [8] of the samples, we also tested the feasibility of our device for the same. In this direction, we measured the height measurement accuracy of our device by placing a mirror (at the sample position) on a piezo-electric stage and measured the movement of the piezo stage at different driving voltages. The efficacy of the height profiling of the CCT system was demonstrated by measuring the height of microfluidic channels on a chip.

## III. RESULTS

The optical response of the aspheric ZnSe lens was first investigated using OpticStudio (Zemax, UK) which is an optical simulation software and then the simulations were experimentally tested using a supercontinuum laser (SCL) source and ZnSe lens combination. The simulation results for 6.35 mm (39469, Edmund Optics, UK), 12.7 mm (39471, Edmund Optics, UK), and 50.8 mm (39495, Edmund Optics, UK) focal length lens at 500 nm, 700 nm, and 950 nm wavelengths are shown in Table 1. From the simulations, it is evident that the lateral resolution, ranging between 0.7 $\mu$m to 1.57 $\mu$m for 500 to 950 nm wavelengths, can be achieved by using a 6.35 mm focal length lens with an imaging range of 1.57 mm. We have defined a parameter, extended depth resolution ratio (EDRR), which is the ratio of the imaging range and average resolution in that region. The EDRR for the 6.35 mm focal length lens comes out to be 1414. The imaging range can be increased to 3.4 mm by using a 12.7 mm focal length lens, and can be increased further up to 19.38 mm by using a 50.8 mm focal length lens, with an EDRR of 2377 and 7814 respectively.

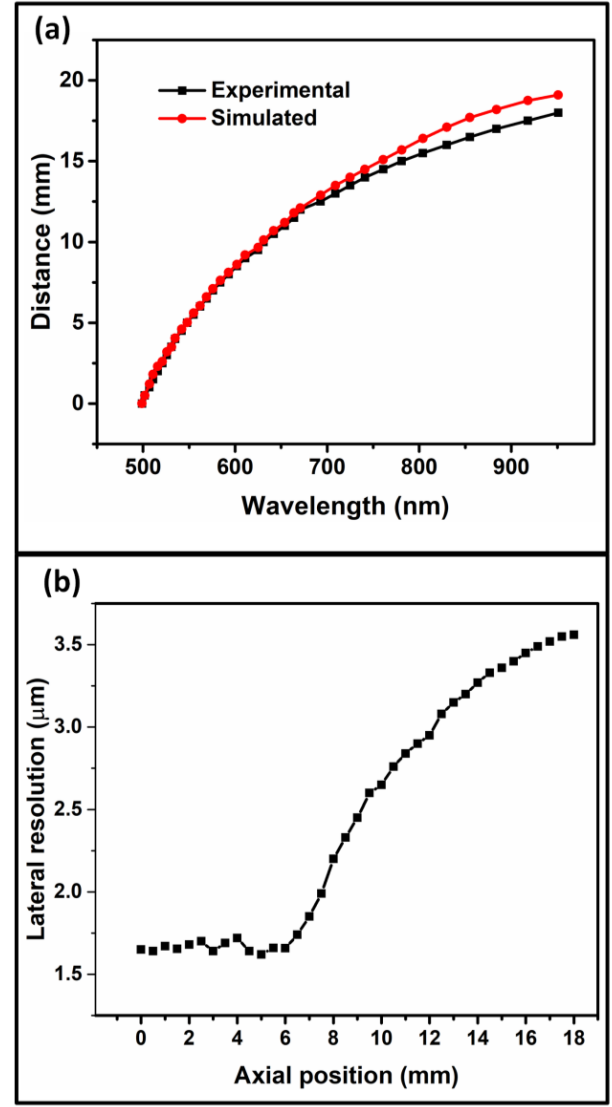

Fig. 2. Characterization of 50.8 mm focal length, ZnSe lens-based CCT system. (a) The graph plotted between the focal distance and wavelength. (b) The experimental lateral resolution as a function of the axial position.

To experimentally demonstrate the capability of ZnSe based CCT system, we chose 6.35 mm and 50.8 mm focal length lenses as one provides better resolution and the other provides a larger imaging range. The object space numerical aperture for 6.35 mm and 50.8 mm focal length lenses at 700 nm wavelength was measured to be 0.52 and 0.16 respectively with an effective focal length of 6.5 mm and 77.9 mm respectively. We characterized the 50.8 mm focal length, ZnSe lens-based CCT system by measuring the imaging range, the lateral resolution, and the axial resolution. The axial focal position of different wavelength components over the wavelength range is shown in Fig. 2 (a). We achieved an imaging range of 18 mm for the wavelength components from 500 nm to 950 nm which is in close agreement with the simulations. In Fig. 2 (b) we show the experimental lateral resolution across the entire imaging range. For the 50.8 mm focal length, ZnSe lens-based CCT system, we could achieve an average lateral resolution of 2.46 μm and average axial resolution of 99.4 μm where the averaging is performed on measurement points separated equally in distance space.

Since for the biological samples, better resolution is more desirable, we designed a CCT system using a 6.35 mm focal length ZnSe lens (39469, Edmund Optics, UK) which was used for imaging of various samples. The system was designed and characterized theoretically and experimentally for different parameters like chromatic axial separation over the bandwidth of the light source, axial, and lateral resolution. The axial separation between various wavelength components as a function of wavelength was calculated theoretically and measured experimentally and is plotted in Fig. 3. As shown in Fig. 3 (a), the wavelengths from 500 nm to 950 nm are axially separated over a distance of 1.55 mm. In Fig. 3 (b) we show the peak position of the reflected signal from a mirror at different distances as a function of camera pixel numbers.

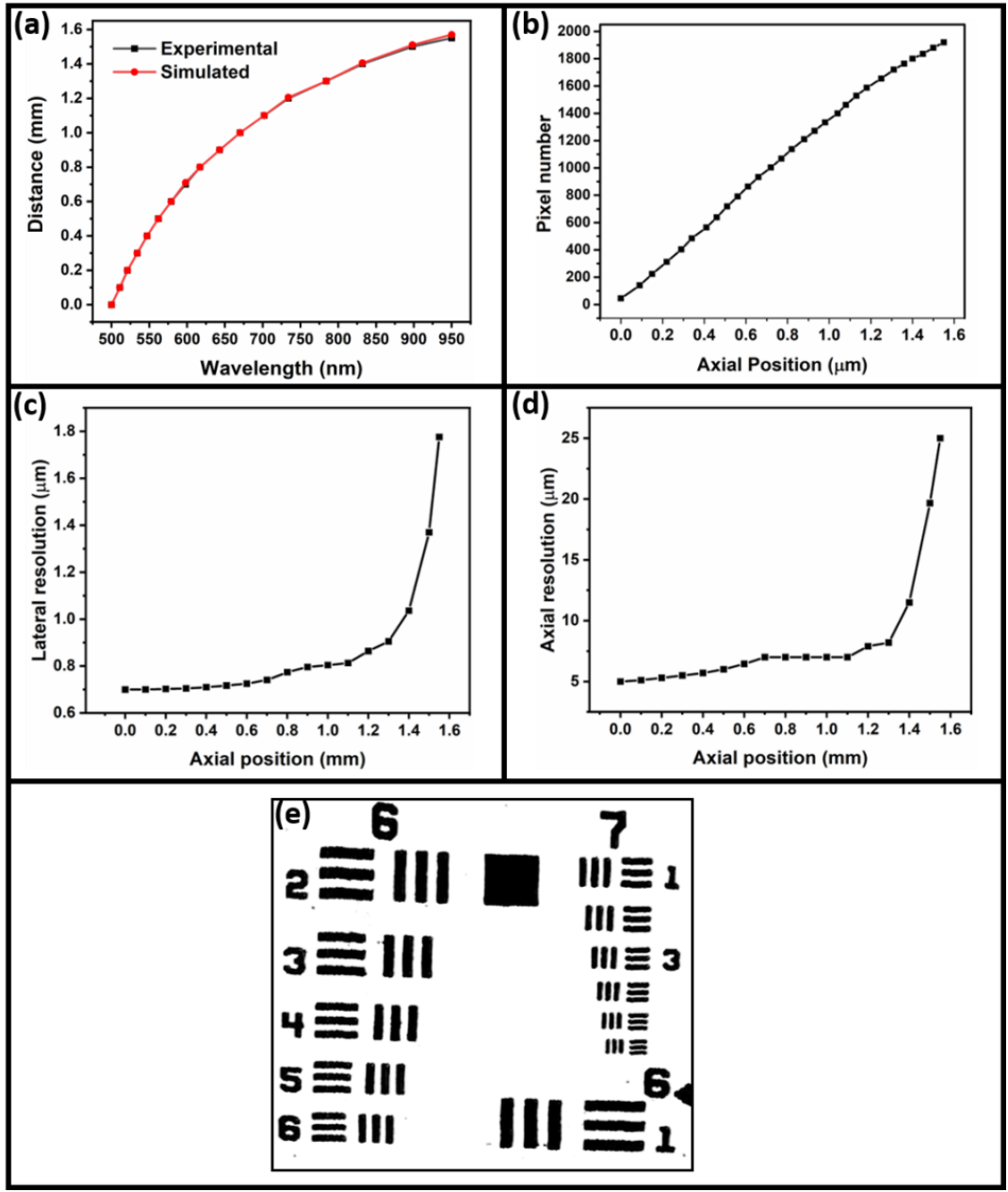

Fig. 3. Characterization of 6.35 mm ZnSe lens based CCT system. (a) The graph plotted between the wavelength and their corresponding focal distance. (b) The peak position of the reflected signal at the camera pixel number with respect to the axial position of the reflecting mirror. (c) The experimental lateral resolution as a function of the axial position. (d) The axial resolution measured as a function of the axial position. (e) Enface image of the USAF resolution target measured at the focal position of 950 nm wavelength.

The lateral resolution for the 6.35 mm ZnSe lens-based CCT device was measured experimentally using the knife-edge technique. The experimentally measured lateral resolution is plotted in Fig. 3 (c). The best lateral resolution was measured to be 0.7 µm at 500 nm wavelength and was maintained below 1 µm up to 735 nm which corresponds to a depth of approximately 1.2 mm. For 950 nm wavelength, the lateral resolution increased to 1.77 µm. The average lateral resolution with the 6.35 mm focal length ZnSe lens was found to be 0.87 µm where the measurement points are equally separated in distance along the axial direction. Further, we imaged the USAF resolution target (R1S1L1N, Thorlabs Inc., USA) at 950 nm wavelength, and could identify all the six elements of group seven as shown in Fig. 3 (e). This corresponds to a lateral resolution better than 2 µm which was maintained for the whole imaging range.

We also measured the axial resolution which determines the optical sectioning capability of the CCT device. The measured axial resolution as a function of the axial position is shown in Fig. 3 (d). Within the wavelength range of 500 nm to 950 nm, the axial resolution varies from 5 µm to 25 µm. The average axial resolution was found to be 8.6 µm over the entire imaging range where data points are equally separated in distance along the axial direction.

To demonstrate the high-resolution depth imaging capability of the CCT device, we performed volumetric imaging of an iron oxide nanoparticles phantom with particle size <1 µm. The volumetric image of the phantom is shown in Fig. 4 (a). The phantom was imaged for a 0.5 mm × 0.5 mm area with a depth of around 1.5 mm (limited by the depth range of our device). The enface images of the iron oxide nanoparticles at depths of 0.35 mm, 0.9 mm, and 1.5 mm are shown in Fig. 4 (b), (c), and (d) respectively. The corresponding X and Y intensity profiles of the visibly smallest oxide nanoparticle are plotted in Fig. 4 (e), (f), and (g) respectively. At a depth of 0.35 mm, the size of the oxide nanoparticle is measured to be better than 1 µm in X-axis and Y-axis. The size of the smallest nanoparticle at a depth of 0.9 mm was also measured to be of the order of 1 µm. The measured size of the nanoparticle increased to ∼ 1.8 µm at a depth of 1.5 mm.

To test the potential of the tomographic system to image biological samples, we imaged a swine cornea *ex-vivo* acquired from a local butcher shop. The imaging of cornea *ex-vivo* did not require any approval from the ethical review board. The volumetric image (0.5mm × 0.5mm × 1.5mm), and the enface images of the cornea (0.5mm × 0.5mm) at three different locations (at the top, the middle and the bottom of the cornea) are shown in Fig. 5 (a), (b), (c), and (d) respectively. From the *enface* images of the cornea at different depths, one can see different cellular features. Around the top of the cornea (Fig. 5(b)), the epithelial cells are seen. In the middle of the cornea (Fig. 5(c)), fibrous structures of the stroma can be seen.

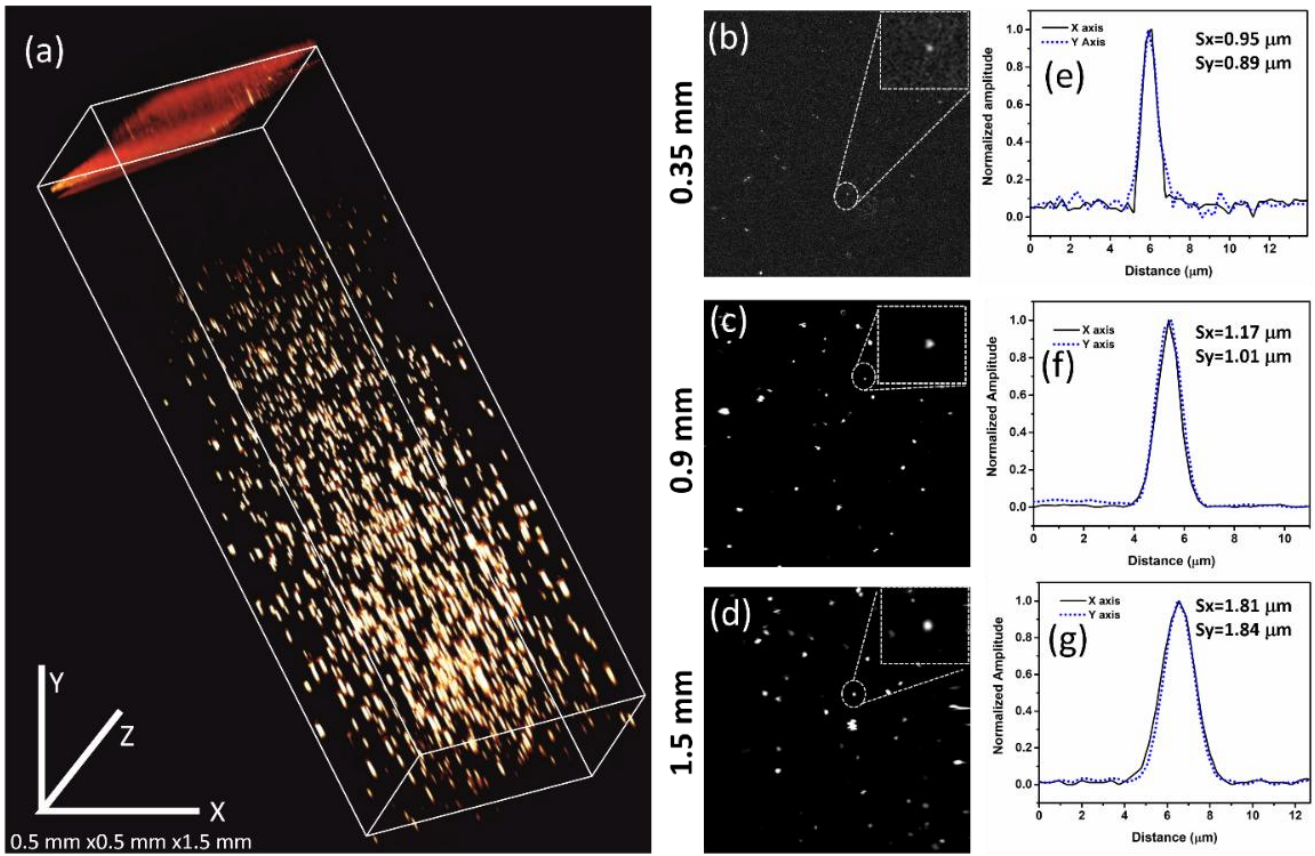

Fig. 4. Volumetric imaging of iron oxide nanoparticle phantom. (a) 3D image of the iron oxide nanoparticle phantom. Enface images of the nanoparticles at the depth of (b) 0.35 mm, (c) 0.9 mm, and (d) 1.5 mm. X and Y axis intensity profile of the smallest nanoparticle at the depth of (e) 0.35 mm, (f) 0.9 mm, and (g) 1.5 mm.

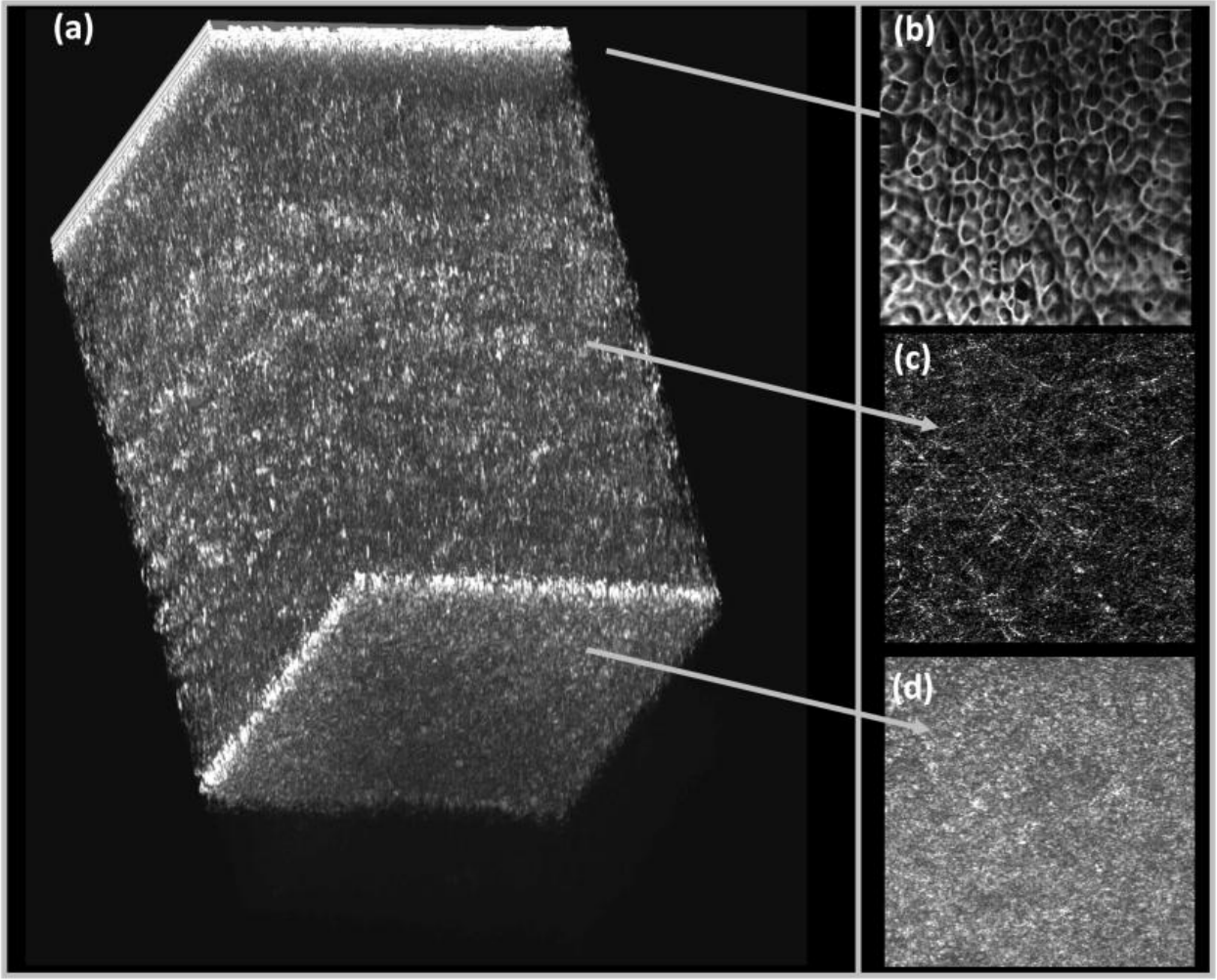

Fig. 5. Confocal images of the cornea tissue. (a) 3D reconstruction of the cornea (0.5mm × 0.5mm × 1.5mm). *Enface* image of the (b) top corneal surface (0.5mm × 0.5 mm) (c) middle of the cornea (0.5mm × 0.5 mm) and (d) below the cornea (0.5mm × 0.5 mm).

An important application of the CCT system is in the field of surface profiling. For this, we measured the minimum height change that one can measure using our device. To do so, we placed a mirror at the focal position of 700 nm wavelength and monitored the peak position of the reflected signal which is shown in Fig. 6 (a). The data was acquired at 51.2 kHz sampling rate. The standard deviation of the peak position was measured to be 14 nm without any averaging, making our device capable of surface profiling with an accuracy of 14 nm. The surface profiling accuracy can be further improved if averaging on consecutive axial scans is performed which however will reduce the imaging speed. Further, we introduced a controlled displacement in the mirror placed at the sample position using a piezo-electric stage. The displacement measured by our device when the piezo stage was driven with a sinusoidal voltage for ~ 50 nm, and ~ 900 nm displacement is shown in Fig. 6 (b), and 6 (c) respectively.

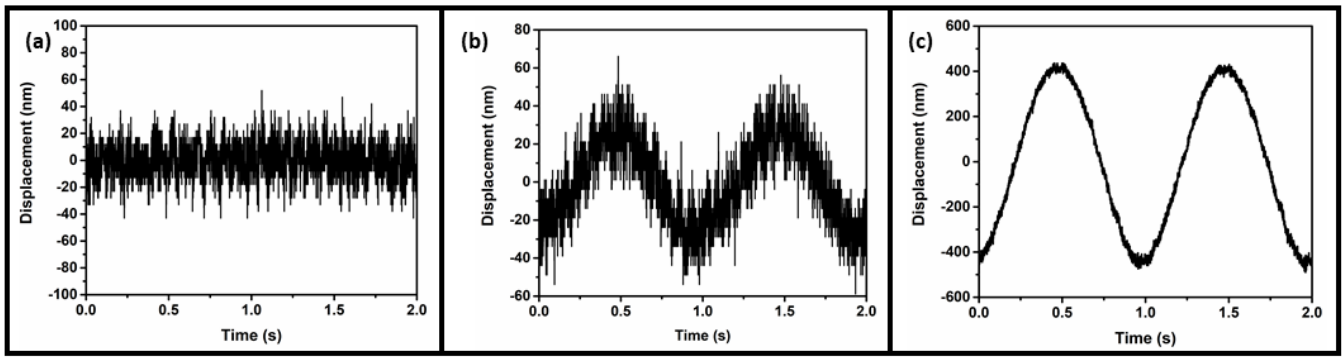

Fig. 6. Height profiling accuracy measurement. (a) Displacement measured for a static mirror which represents the noise of the system in terms of displacement measurement. The measured displacement of the mirror placed on a piezo-electric stage when driven with a sinusoidal voltage for (b) ~ 50 nm, and (c) ~ 900 nm displacement.

Furthermore, we performed the surface profiling of an in-house fabricated microfluidic channel to demonstrate the height profiling feasibility of our device. The microfluidic channel was designed with a channel of varying width but of a constant height of 30 micrometers. In Fig. 7, we show the volumetric reconstruction of the microfluidic channel (Fig. 7a) and a B-Scan (Fig. 7 (b)) to show the height profile of one of the narrow channels. The intensity profile at the location of asterisks within Fig. 7 (b) as a function of distance is shown in Fig. 7 (c). The measured dimensions of the microfluidic channel match very well with the designed dimensions.

## IV. DISCUSSION AND CONCLUSION

Conventionally, lenses based on materials such as fused silica and quartz are used in the visible-NIR range due to their excellent optical quality and high transmission ~90%. These materials have an Abbe number of around 50. Zinc selenide has an Abbe number ~8 for visible wavelengths, making it, to the best of our knowledge, the lowest Abbe number material which is commercially available for the visible/NIR wavelengths. However, ZnSe is rarely used for visible wavelengths because of its low transmission; around 50% for the 550 nm and 60% for the 600 nm and above. Even though ZnSe has a low transmission in the visible and NIR range, it makes a great candidate for chromatic confocal tomography because of its high chromatic dispersion. In this work, we have utilized the chromatic dispersion property of a single ZnSe lens (50.8 mm focal length) to realize a CCT system with an imaging range of 1.8 cm with an average lateral resolution of 2.46 µm. Using a 6.37 mm focal length ZnSe lens, we could improve the average lateral resolution to 0.87 µm with an imaging range of 1.55 mm.

Besides ZnSe, Gallium phosphide (GaP) is another material that can be used for extended imaging range CCT device as it also has a low Abbe number ~6 in the visible range. However, it has not been used for commercial lens fabrication due to the release of dangerous substances during the fabrication process [25]. Furthermore, although ZnSe lenses are commercially available, care should be taken while handling these lenses as ZnSe material is also known to be a hazardous material.

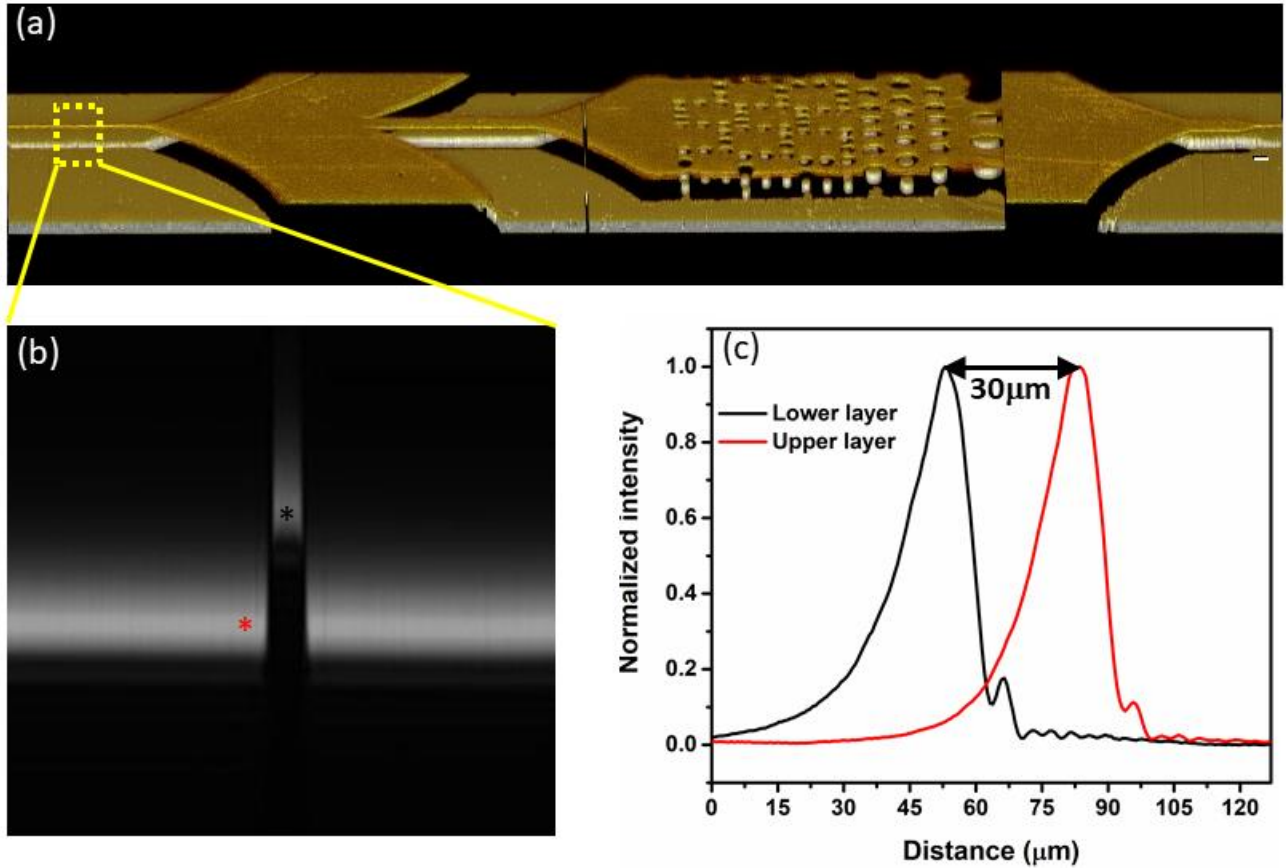

Fig. 7. Height profile measurement. (a) Volumetric 3D reconstruction of the microfluidic channel chip (b) B-Scan of the microfluidic channel at the position indicated by the box. (c) Intensity profile of the two layers as a function of distance at locations indicated by the asterisks (black and red) in (b).

In general, the aspheric lenses are corrected for the spherical aberrations but only for a certain wavelength. As we move away from the spherically corrected wavelength, the aberrations increase and reduce the image quality. The ZnSe lens used in our system was designed for 10 μm wavelength and therefore spherically corrected for wavelengths around this wavelength only. If this lens is used for focusing a collimated NIR/IR beam of light, the spherical aberrations would deteriorate the focal spot significantly. Therefore we used a diverging beam from the output of the optical fiber which allowed us to minimize the spherical aberration for the visible-NIR wavelengths. Depending on the NA of the optical fiber, the distance between the fiber exit and the ZnSe lens can be adjusted to minimize the spherical aberrations for a certain wavelength. We used this methodology to design our system and the distance between the fiber tip and the front surface of the lens was adjusted to minimize the spherical aberrations for 700 nm. The spherical aberrations for the rest of the spectrum were not fully compensated but it still allowed us to achieve an average lateral resolution of 0.87 μm and 2.46 μm for 6.35 mm and 50.8 mm focal length lenses respectively.

Similar to other point scanning confocal microscopes, our device is also limited in terms of image acquisition speed. The current system allows us to acquire full volumetric data (512 × 512 × 2048 pixels) in approximately 5.12 seconds which can be improved further by using a fast scanner and camera combination.

One important aspect that we should clarify is that although our approach provides an extended imaging range, the penetration depth in the sample is still limited and will be dictated by the sample properties such as absorption and scattering. For instance, the penetration depth of the visible wavelengths in a biological sample is typically a few hundred micrometers and it can be argued that the extended imaging range that our approach provided is not usable. This would be true if the biological samples were perfectly flat, which is not always the case. For example in a coronary artery with calcified plaque [26], the surface height can change by a few millimeters because of the calcification below the surface. Also, the tissue surface of the gastrointestinal tract is highly irregular [27] and for imaging of such tissue, not only penetration depth but also an extended imaging range is required.

Celluar level three-dimensional imaging of biological samples is an obvious application of a confocal microscope but another important application for a chromatic confocal microscope is height profiling. Using a 6.37 mm focal length ZnSe lens, our CCT device would allow one to measure the height profile up to 1.55 mm with a height accuracy of 14 nm and an average lateral resolution of 0.87 μm. The height measurement accuracy can be improved further by averaging the measurements but it comes at the expense of acquisition time. Several such devices are commercially available where chromatic aberrations of standard lenses are used for the axial separation of the spectral components. The reflected light from the sample is acquired in the confocal configuration and the peak of the reflected spectrum is used to measure the height profile of the sample. As an example, one of the commercially available systems (IFS2407-0,1, from Micro-Epsilon) which has a spatial resolution of 3 micrometers (comparable to our system), provides an imaging range of 100 micrometers. Using a 50.8 mm focal length ZnSe lens, our CCT device could achieve an imaging range of 1.8 cm, which is 180 times greater than the commercially available device. Our CCT system design not only allowed us to use the commercially available lenses but also the fact that only a single lens is required for 3D imaging, makes the system very attractive for both commercial and biological applications.

In conclusion, we have designed and developed a single ZnSe based chromatic confocal tomography system with an imaging range of 1.8 cm and with an average spatial resolution of 2.4 μm. For high-resolution biological applications, using a shorter focal length lens the average spatial resolution can be improved further to 0.87 μm with an imaging range of 1.55 mm. Our device is capable of imaging biological samples in 3D at cellular resolution and can perform surface profiling of the biological as well as industrial samples with an unmatched combination of spatial resolution and imaging range. We believe that this advancement will find several applications in the biomedical and industrial fields.

## APPENDIX AND THE USE OF SUPPLEMENTAL FILES

See Supplement 1 for supporting content.

## ACKNOWLEDGMENT

Dr. Gargi Sharma would like to thank Prof. Jochen Guck and Dr. Girardo for providing microfluidic channel for height profiling measurements.

## REFERENCES

[1]    D. Huang, E. A. Swanson, C. P. Lin, J. S. Schuman, W. G. Stinson, W. Chang, *et al.*, "Optical coherence tomography," *Science*, vol. 254, p. 1178, 1991.

[2]    S. W. Paddock, "Confocal Laser Scanning Microscopy," *BioTechniques*, vol. 27, pp. 992-1004, 1999/11/01 1999.

[3]    K. Shi, P. Li, S. Yin, and Z. Liu, "Chromatic confocal microscopy using supercontinuum light," *Optics Express*, vol. 12, pp. 2096-2101, 2004/05/17 2004.

[4]    L. Liu, J. A. Gardecki, S. K. Nadkarni, J. D. Toussaint, Y. Yagi, B. E. Bouma, *et al.*, "Imaging the subcellular structure of human coronary atherosclerosis using micro–optical coherence tomography," *Nature Medicine*, vol. 17, pp. 1010-1014, 2011/08/01 2011.

[5]    J.-A. Conchello and J. W. Lichtman, "Optical sectioning microscopy," *Nature Methods*, vol. 2, pp. 920-931, 2005/12/01 2005.

[6]    E. Wang, C. M. Babbey, and K. W. Dunn, "Performance comparison between the high-speed Yokogawa spinning disc confocal system and single-point scanning confocal systems," *Journal of Microscopy*, vol. 218, pp. 148-159, 2005/05/01 2005.

[7]    S. W. Paddock, "Principles and practices of laser scanning confocal microscopy," *Molecular Biotechnology*, vol. 16, pp. 127-149, 2000/10/01 2000.

[8] H. J. Tiziani and H. M. Uhde, "Three-dimensional image sensing by chromatic confocal microscopy," *Applied Optics,* vol. 33, pp. 1838-1843, 1994/04/01 1994.

[9] H. J. Tiziani, R. Achi, and R. N. Krämer, "Chromatic confocal microscopy with microlenses," *Journal of Modern Optics,* vol. 43, pp. 155-163, 1996/01/01 1996.

[10] S. Cha, P. C. Lin, L. Zhu, P.-C. Sun, and Y. Fainman, "Nontranslational three-dimensional profilometry by chromatic confocal microscopy with dynamically configurable micromirror scanning," *Applied Optics,* vol. 39, pp. 2605-2613, 2000/06/01 2000.

[11] C. Olsovsky, R. Shelton, O. Carrasco-Zevallos, B. E. Applegate, and K. C. Maitland, "Chromatic confocal microscopy for multi-depth imaging of epithelial tissue," *Biomedical Optics Express,* vol. 4, pp. 732-740, 2013/05/01 2013.

[12] S. Li and R. Liang, "DMD-based three-dimensional chromatic confocal microscopy," *Applied Optics,* vol. 59, pp. 4349-4356, 2020/05/10 2020.

[13] C. Chen, W. Song, J.-W. Chen, J.-H. Wang, Y. H. Chen, B. Xu, *et al.*, "Spectral tomographic imaging with aplanatic metalens," *Light: Science & Applications,* vol. 8, p. 99, 2019/11/06 2019.

[14] P. C. Lin, P.-C. Sun, L. Zhu, and Y. Fainman, "Single-shot depth-section imaging through chromatic slit-scan confocal microscopy," *Applied Optics,* vol. 37, pp. 6764-6770, 1998/10/01 1998.

[15] S. L. Dobson, P.-c. Sun, and Y. Fainman, "Diffractive lenses for chromatic confocal imaging," *Applied Optics,* vol. 36, pp. 4744-4748, 1997/07/10 1997.

[16] W. Lyda, M. Gronle, D. Fleischle, F. Mauch, and W. Osten, "Advantages of chromatic-confocal spectral interferometry in comparison to chromatic confocal microscopy," *Measurement Science and Technology,* vol. 23, p. 054009, 2012/03/22 2012.

[17] B. S. Chun, K. Kim, and D. Gweon, "Three-dimensional surface profile measurement using a beam scanning chromatic confocal microscope," *Review of Scientific Instruments,* vol. 80, p. 073706, 2009/07/01 2009.

[18] T. Kim, S. H. Kim, D. Do, H. Yoo, and D. Gweon, "Chromatic confocal microscopy with a novel wavelength detection method using transmittance," *Optics Express,* vol. 21, pp. 6286-6294, 2013/03/11 2013.

[19] J. G. R, J. Meneses, G. Tribillon, T. Gharbi, and A. Plata, "Chromatic confocal microscopy by means of continuum light generated through a standard single-mode fibre," *Journal of Optics A: Pure and Applied Optics,* vol. 6, pp. 544-548, 2004/04/24 2004.

[20] J. Garzón, T. Gharbi, and J. Meneses, "Real time determination of the optical thickness and topography of tissues by chromatic confocal microscopy," *Journal of Optics A: Pure and Applied Optics,* vol. 10, p. 104028, 2008/09/02 2008.

[21] D. Luo, C. Kuang, and X. Liu, "Fiber-based chromatic confocal microscope with Gaussian fitting method," *Optics & Laser Technology,* vol. 44, pp. 788-793, 2012/06/01 2012.

[22] A. Gerger, S. Koller, T. Kern, C. Massone, K. Steiger, E. Richtig, *et al.*, "Diagnostic Applicability of In Vivo Confocal Laser Scanning Microscopy in Melanocytic Skin Tumors," *Journal of Investigative Dermatology,* vol. 124, pp. 493-498, 2005/03/01 2005.

[23] R. Alvarez-Román, A. Naik, Y. N. Kalia, H. Fessi, and R. H. Guy, "Visualization of skin penetration using confocal laser scanning microscopy," *European Journal of Pharmaceutics and Biopharmaceutics,* vol. 58, pp. 301-316, 2004/09/01 2004.

[24] J. TrÄGÅRdh, K. Macrae, C. Travis, R. Amor, G. Norris, S. H. Wilson, *et al.*, "A simple but precise method for quantitative measurement of the quality of the laser focus in a scanning optical microscope," *Journal of Microscopy,* vol. 259, pp. 66-73, 2015/07/01 2015.

[25] J. Václavík and D. Vápenka, "Gallium Phosphide as a material for visible and infrared optics," *EPJ Web of Conferences,* vol. 48, // 2013.

[26] H. Mori, S. Torii, M. Kutyna, A. Sakamoto, A. V. Finn, and R. Virmani, "Coronary Artery Calcification and its Progression: What Does it Really Mean?," *JACC: Cardiovascular Imaging,* vol. 11, pp. 127-142, 2018/01/01 2018.

[27] M. J. Gora, J. S. Sauk, R. W. Carruth, K. A. Gallagher, M. J. Suter, N. S. Nishioka, *et al.*, "Tethered capsule endomicroscopy enables less invasive imaging of gastrointestinal tract microstructure," *Nature Medicine,* vol. 19, pp. 238-240, 2013/02/01 2013.